\documentclass[nofootinbib,noshowpacs,noshowkeys,a4paper,amsmath,amssymb,nopacs,12pt]{revtex4}

\begin{document}

\title{Opposite parity fermion mixing and baryons $1/2^{\pm}$}

\author{A.E. Kaloshin}
\affiliation{Irkutsk State University}

\author{E.A. Kobeleva}
\affiliation{Irkutsk State University}

\author{V.P. Lomov}
\affiliation{The Institute for System Dynamics and Control Theory of SB RAS}

\begin{abstract}
  We develop a variant of $K$-matrix, which includes the effect of opposite parity fermions
  (OPF) mixing, and apply it for description of $\pi N$ partial waves $S_{11}$ and
  $P_{11}$. OPF-mixing leads to appearance of negative energy poles in $K$-matrix and
  restoration of MacDowell symmetry, relating two partial waves. Joint analysis of PWA results
  for $S_{11}$ and $P_{11}$ confirms significance of this effect.
\end{abstract}

\maketitle

\section{Introduction}
\label{sec:intro}

For fermions there exists a non-standard mixing, when fermion fields with opposite parities are
mixing at loop level while parity is conserved in vertex (shortly OPF-mixing):%%%
{\par\centering
  \includegraphics{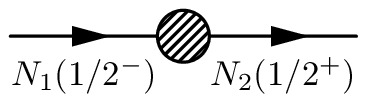}\par
}
\noindent
It is possible because fermion and antifermion have different parities.  This effect was
investigated in detail in \cite{Kaloshin:2010jj} and was applied to $\pi N$ scattering, where it
leads to relation between two partial waves.  In \cite{Kaloshin:2010jj} was found the simplest
physical example of manifestation of this effect: the partial waves $P_{13}$ and $D_{13}$, where
baryons $J=3/2^\pm$ are produced. The OPF-mixing effect is identified in the partial wave
$P_{13}$ as rather specific interference of resonance with background generated by resonance
state in $D_{13}$ wave. The above-mentioned relation between partial waves influences mainly on
a wave with lower orbital momentum and it is used as additional source of information about
structure of wave with higher $l$.

Another physical example, where OPF-mixing may be essential, is related with the partial waves
$S_{11}$ and $P_{11}$, where resonances $J^{P}=1/2^{\pm}, I=1/2$ are produced. Most interesting
object here is the Roper resonance $N(1440)$, which has some unusual properties and problems
with quark-models identification, see,
e.g. \cite{Krehl:1999km,Batinic:1995kr,Glozman:1995fu,Capstick:2000qj,Mathur:2003zf,Dillig:2004rh,Sarantsev:2007bk,JuliaDiaz:2006av,Roberts:2007ji}. However,
in presence of several resonance states the approach of \cite{Kaloshin:2010jj}, that uses a
matrix propagator, becomes too cumbersome.  Alternatively, for description of OPF-mixing one can
use the $K$-matrix approach, which works for any number of states and channels.

In this paper we develop the $K$-matrix approach for $\pi N$ partial amplitudes with accounting
of the OPF-mixing effect and apply it for description of $S_{11}$ and $P_{11}$ partial
waves. Most serious changing as compared with its standard form is the appearance of negative
energy poles in $K$-matrix. If, besides, we use QFT to calculate tree amplitudes
(i.e. $K$-matrix), starting from effective Lagrangians, we obtain the partial amplitudes $\pi
N\to\pi N$ satisfying the MacDowell symmetry condition:
\begin{equation}\label{eq:macdowell}
  f_{l,+}(W)=-f_{l+1,-}(-W), 
\end{equation}
which was obtained \cite{MacDowell:1959zza} from general analytic properties of amplitudes.

We use the obtained $K$-matrix to describe results of partial wave analysis for $S_{11}$ and
$P_{11}$ amplitudes. The main purpose is to see the manifestation of OPF-mixing and it naturally
leads to joint fitting of these two waves.

\section{Mixing of fermions with opposite parities and $K$-matrix}
\label{sec:mixing-and-K-matrix}

We need to discuss the effect of OPF-mixing in amplitudes of $\pi N$ scattering and its
implementation in framework of $K$-matrix description. For a first step one may restrict oneself
by a simplified case: two resonance states and two channels. Let us write down the effective
Lagrangians $\pi N N'$ without derivatives and conserving the parity:
\begin{alignat}{2}
  \label{eq:eff-lagr-m}
  \Lagr_{\text{int}} &= g_{1}\bar{N}_{1}(x)N(x)\phi(x)+\hc, &\quad& \text{for }
  J^{P}(N_{1})=1/2^{-},\\
  \label{eq:eff-lagr-p}
  \Lagr_{\text{int}} &= \imath g_{2} \bar{N}_{2}(x)\gamma^{5}N(x)\phi(x)+\hc, && \text{for }
  J^{P}(N_{2})=1/2^{+}.
\end{alignat}

Let us consider two baryon states of opposite parities with masses $m_{1}$ ($J^{P}=1/2^{-}$),
$m_{2}$ ($J^{P}=1/2^{+}$) and two intermediate states $\pi N$, $\eta N$. Using the effective
Lagrangians we can calculate contributions of states $N_{1}$, $N_{2}$ to partial waves at tree
level (see details in \cite{Kaloshin:2010jj}) for $s$-wave amplitudes:
\begin{equation}\label{eq:pw-s-tree}
  \begin{split}
    f^{\text{tree}}_{s,+}(\pi N\to\pi N) &= -\frac{(E^{(\pi)}_{N}+m_{N})}{8\pi W}
      \bigg(\frac{g_{1,\pi}^{2}}{W-m_{1}}+\frac{g_{2,\pi}^{2}}{W+m_{2}}\bigg),\\
    f^{\text{tree}}_{s,+}(\pi N\to\eta N) &=
      -\frac{\sqrt{\big(E^{(\pi)}_{N}+m_{N}\big)\big(E^{(\eta)}_{N}+m_{N}\big)}}{8\pi W}
      \bigg(\frac{g_{1,\pi}g_{1,\eta}}{W-m_{1}}+\frac{g_{2,\pi}g_{2,\eta}}{W+m_{2}}\bigg),\\
    f^{\text{tree}}_{s,+}(\eta N\to\eta N) &= -\frac{(E^{(\eta)}_{N}+m_{N})}{8\pi W}
      \bigg(\frac{g_{1,\eta}^{2}}{W-m_{1}}+\frac{g_{2,\eta}^{2}}{W+m_{2}}\bigg)
  \end{split}
\end{equation}
and for $p$-wave amplitudes:
\begin{equation}\label{eq:pw-p-tree}
  \begin{split}
    f^{\text{tree}}_{p,-}(\pi N\to\pi N) &= \frac{(E^{(\pi)}_{N} - m_{N})}{8\pi W}
      \bigg(\frac{g_{1,\pi}^{2}}{-W-m_{1}}+\frac{g_{2,\pi}^{2}}{-W+m_{2}}\bigg),\\
    f^{\text{tree}}_{p,-}(\pi N\to\eta N) &=
      \frac{\sqrt{\big(E^{(\pi)}_{N} - m_{N}\big)\big(E^{(\eta)}_{N} - m_{N}\big)}}{8\pi W}
      \bigg(\frac{g_{1,\pi}g_{1,\eta}}{-W-m_{1}}+\frac{g_{2,\pi}g_{2,\eta}}{-W+m_{2}}\bigg),\\
    f^{\text{tree}}_{p,-}(\eta N\to\eta N) &= \frac{(E^{(\eta)}_{N} - m_{N})}{8\pi W}
      \bigg(\frac{g_{1,\eta}^{2}}{-W-m_{1}}+\frac{g_{2,\eta}^{2}}{-W+m_{2}}\bigg).
  \end{split}
\end{equation}
Here $W=\sqrt{s}$ is the total CMS energy and $E^{(\pi)}_{N}$ $\big( E^{(\eta)}_{N}\big)$ is
nucleon CMS energy of system $\pi N$ $\big(\eta N\big)$
\begin{equation}\label{eq:1}
  E^{(\pi)}_{N}=\frac{W^{2}+m_{N}^{2}-m_{\pi}^{2}}{2W}.
\end{equation}
We introduced here short notation for coupling constants, e.g. $g_{1,\pi}=g_{N_{1}N\pi}$.

The tree amplitudes \eqref{eq:pw-s-tree}--\eqref{eq:pw-p-tree} contain poles with both positive
and negative energy, originated from propagators of $N_{1}$ and $N_{2}$ fields of opposite
parities. Accounting the loop transitions results in dressing of states and also in mixing of
these two fields.

Note that $W \to -W$ replacement gives
\begin{equation}\label{eq:2}
  E^{(\pi)}_{N}+m_{N}\to-\big(E^{(\pi)}_{N}-m_{N}\big),
\end{equation}
so tree amplitudes \eqref{eq:pw-s-tree}--\eqref{eq:pw-p-tree} possess the MacDowell symmetry
property \cite{MacDowell:1959zza}
\begin{equation}\label{eq:macdowell2}
  f_{p,-}(W)=-f_{s,+}(-W).
\end{equation}

In $K$-matrix representation for partial amplitudes
\begin{equation}\label{eq:K-matrix-rep}
  f=K \big(1-\imath PK \big)^{-1},
\end{equation}
diagonal matrix $\imath P$, constructed from CMS momenta, originates from imaginary part of a
loop. Therefore, $K$-matrix here is simply a matrix of tree amplitudes that should be identified
with amplitudes \eqref{eq:pw-s-tree},\eqref{eq:pw-p-tree}.

As the result we come to representation of partial amplitudes for $s$- and $p$-waves
\begin{equation}\label{eq:K-matrix-pw-s-p}
  f_{s}(W)=K_{s}(W)\big(1-\imath PK_{s}(W) \big)^{-1},\quad
  f_{p}(W)=K_{p}(W)\big(1-\imath PK_{p}(W) \big)^{-1},
\end{equation}
where the matrices $K_{s}$, $K_{p}$ (i.e. tree amplitudes
\eqref{eq:pw-s-tree},\eqref{eq:pw-p-tree}), may be written in factorized form \footnote{Similar
  $K$-matrix have been used for a long time in $\pi N$ phenomenology, see, e.g. \cite{Arndt:1985vj},
  but with other phase-space factors.}
\begin{equation}\label{eq:4}
  K_{s}=-\frac{1}{8\pi}\rho_{s}\hat{K}_{s}\rho_{s},\quad
  K_{p}= \frac{1}{8\pi}\rho_{p}\hat{K}_{p}\rho_{p}.
\end{equation}
Here $\rho_{s}$, $\rho_{p}$ are
\begin{equation}\label{eq:5}
  \rho_{s}(W)=
  \begin{pmatrix}
    \sqrt{\dfrac{E^{(\pi)}_{N}+m_{N}}{W}}, & 0 \\
    0, & \sqrt{\dfrac{E^{(\eta)}_{N}+m_{N}}{W}}
  \end{pmatrix},\quad
  \rho_{p}(W)=
  \begin{pmatrix}
    \sqrt{\dfrac{E^{(\pi)}_{N}-m_{N}}{W}}, & 0\\
    0, & \sqrt{\dfrac{E^{(\eta)}_{N}-m_{N}}{W}}
  \end{pmatrix},
\end{equation}
and matrix $P$ consists of CMS momenta as analytic functions of $W$. In this case "primitive"
$K$-matrices contain poles with both positive and negative energy
\begin{align}
  \label{eq:K-bar-s-prim}
  \hat{K}_{s}(W)&=
  \begin{pmatrix}
    \dfrac{g_{1,\pi}^{2}}{W-m_{1}}+\dfrac{g_{2,\pi}^{2}}{W+m_{2}}, &
    \dfrac{g_{1,\pi}g_{2,\eta}}{W-m_{1}}+\dfrac{g_{2,\pi}g_{2,\eta}}{W+m_{2}}\\
    \dfrac{g_{1,\pi}g_{2,\eta}}{W-m_{1}}+\dfrac{g_{2,\pi}g_{2,\eta}}{W+m_{2}},&
    \dfrac{g_{1,\eta}^{2}}{W-m_{1}}+\dfrac{g_{2,\eta}^{2}}{W+m_{2}}\\
  \end{pmatrix},\\
  \label{eq:K-bar-p-prim}
  \hat{K}_{p}(W)=\hat{K}_{s}(-W)&=
  \begin{pmatrix}
    \dfrac{g_{1,\pi}^{2}}{-W-m_{1}}+\dfrac{g_{2,\pi}^{2}}{-W+m_{2}}, &
    \dfrac{g_{1,\pi}g_{2,\eta}}{-W-m_{1}}+\dfrac{g_{2,\pi}g_{2,\eta}}{-W+m_{2}}\\
    \dfrac{g_{1,\pi}g_{2,\eta}}{-W-m_{1}}+\dfrac{g_{2,\pi}g_{2,\eta}}{-W+m_{2}},&
    \dfrac{g_{1,\eta}^{2}}{-W-m_{1}}+\dfrac{g_{2,\eta}^{2}}{-W+m_{2}}\\
  \end{pmatrix}.
\end{align}
Recall that $m_{1}$ is mass of $J^{P}=1/2^{-}$ state and $m_{2}$ is mass of $J^{P}=1/2^{+}$
one. Generalization of this construction for the case of more channels and states is obvious.

Since CMS momenta have the property $P(-W)=-P(W)$, the MacDowell symmetry property
\eqref{eq:macdowell2} is extended from tree amplitudes to unitarized $K$-matrix ones
\eqref{eq:K-matrix-pw-s-p}. Note that our $K$-matrix amplitudes \eqref{eq:K-matrix-pw-s-p} may
be rewritten in other form, close to the one used in \cite{Arndt:1985vj}
\begin{equation}\label{eq:6}
  \begin{split}
    f_{s}(W) &= -\frac{1}{8\pi}\rho_{s}\hat{K}_{s}
      \big[1+\imath\rho_{s}P\rho_{s}\hat{K}_{s}(W)/(8\pi)\big]^{-1}\rho_{s},\\
    f_{p}(W) &= \phantom{-}\frac{1}{8\pi}\rho_{p}\hat{K}_{p}
      \big[1-\imath\rho_{p}P\rho_{p}\hat{K}_{p}(W)/(8\pi) \big]^{-1}\rho_{p}.
  \end{split}
\end{equation}

Following a common sense one can expect that presence of negative energy pole, for example, in
elastic $\pi N$ amplitude should give a negligible effect in physical energy region. However,
this is not true if corresponding coupling constant is large
$\abs{g_{2,\pi}}\gg\abs{g_{1,\pi}}$. To see the reason of this ratio, one can compare decay widths
of $s$- and $p$-states
\begin{equation}\label{eq:7}
  \Gamma(N_{1}\to\pi N)=g_{N_{1}\pi N}^{2}\Phi_{s},\quad
  \Gamma(N_{2}\to\pi N)=g_{N_{2}\pi N}^{2}\Phi_{p},
\end{equation}
where $\Phi_{s}$, $\Phi_{p}$ are corresponding phase volumes. For resonance states not far from
threshold, with masses, e.g. $1.5$--$1.7$ GeV, phase volumes differ greatly,
$\Phi_{s}\gg\Phi_{p}$. If both resonances have typical hadronic width $\Gamma\sim100$ MeV, then
coupling constants differ dramatically too, $\abs{g_{N_{2}\pi N}}\gg\abs{g_{N_{1}\pi N}}$. This
inequality will result in increasing of background contribution to $s$-wave and on the other
hand in suppressing of background in $p$-wave. As a result, OPF-mixing leads to relation between
two partial wave of $\pi N$ scattering, but this connection mainly influences on amplitude with
lower orbital number.

Above we use the simplest effective Lagrangians \eqref{eq:eff-lagr-m}--\eqref{eq:eff-lagr-p} to
derive tree amplitudes. However, it is well-known, that spontaneous breaking of chiral symmetry
requires pion field to appear in Lagrangian only through derivative
\begin{equation}\label{eq:eff-lagr-m2}
  \Lagr_{\text{int}}=f_{2}\bar{N}_{2}(x)\gamma^{5}\gamma^{\mu}N(x)\partial_{\mu}\phi(x)+\hc,\quad
  J^{P}=1/2^{+},\quad
  f_{2}=\frac{g_{2}}{m_{2}+m_{N}}.
\end{equation}
It is not difficult to understand how inclusion of derivative changes tree amplitudes and, hence
$K$-matrix. Pole contribution $\pi(k_{1})N(p_{1})\to N_{2}(p)\to\pi(k_{2})N(p_{2})$ in that case
takes the form:
\begin{equation}\label{eq:8}
  T=f_{2}^{2}\bar{u}(p_{2})\gamma^{5}\hat{k}_{2}\frac{1}{\hat{p}-M}\gamma^{5} \hat{k}_{1}u(p_{1}).
\end{equation}
With use of equations of motion, we see that inclusion of derivative at vertex leads to the
following modification of resonance contribution
\begin{equation}\label{eq:3}
  g_{2}^{2}\frac{1}{\hat{p}-M}\to f_{2}^{2}(\hat{p}+m_{N})\frac{1}{\hat{p}-M}(\hat{p}+m_{N}).
\end{equation}
Separation of the positive and negative energy poles is performed with the off-shell projector
operators $\Lambda^{\pm}=1/2\big(1\pm \hat{p}/W\big)$
\begin{equation}\label{eq:9}
  f_{2}^{2}(\hat{p}+m_{N})\frac{1}{\hat{p}-m_{N}}(\hat{p}+m_{N})=
  \Lambda^{+}\frac{f_{2}^{2}(W+m_{N})^{2}}{W-M}+\Lambda^{-}\frac{f_{2}^{2}(W-m_{N})^{2}}{-W-M},
\end{equation}
where the first term gives contribution to $p$-wave and second one to $s$-wave. Modification of the
pole contributions in "primitive" $K$-matrices \eqref{eq:K-bar-s-prim}--\eqref{eq:K-bar-p-prim} is
evident \footnote{It is not difficult to understand that this rule holds for resonance
  contribution of any parity $J^{P}=1/2^{\pm}$.}
\begin{alignat}{2}
  \label{eq:10}
  g_{2}^{2} &\to f_{2}^{2}(W-m_{N})^{2}, &\quad&\text{for $s$-wave,}\\
  \label{eq:11}
  g_{2}^{2} &\to f_{2}^{2}(W+m_{N})^{2}, &&\text{for $p$-wave.}
\end{alignat}
One can expect that the inclusion of derivatives most strongly affects on threshold properties of $s$-wave due to dumping
factor \footnote{As for $J^{P}=1/2^{-}$ baryons, the presence of derivative in vertex leads to
  contradiction with data near threshold in $S_{11}$ due to factor $(W-m_{N})^2$ in
  \eqref{eq:10}, see Fig.~\ref{fig:5}.} $(W-m_{N})^{2}$.

\section{Partial amplitudes of $\pi N$ scattering}
\label{sec:pw-piN}

We will use the described above $K$-matrix for description of partial waves $S_{11}$ and $P_{11}$ of $\pi N$ scattering in the energy region $W<2$ GeV. Following
\cite{Batinic:1995kr,Ceci:2008zz,Ceci:2006zw} we will use three channels of reaction: $\pi N$,
$\eta N$ and $\sigma N$, where the last is "effective" channel, imitating different $\pi\pi N$
states. "Primitive" $K$-matrices have a form \eqref{eq:K-bar-s-prim}--\eqref{eq:K-bar-p-prim} but
can contain several $J^{P}=1/2^{+}$ and $J^{P}=1/2^{-}$ states.

\begin{figure}[!h]
  \centering
  \includegraphics[scale=0.8]{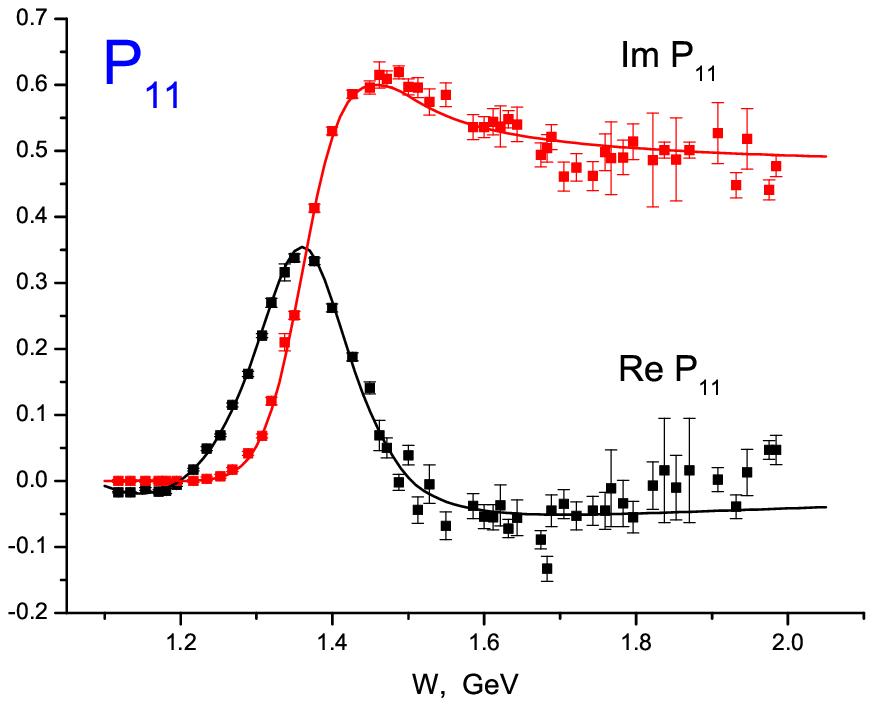}
  \includegraphics[scale=0.8]{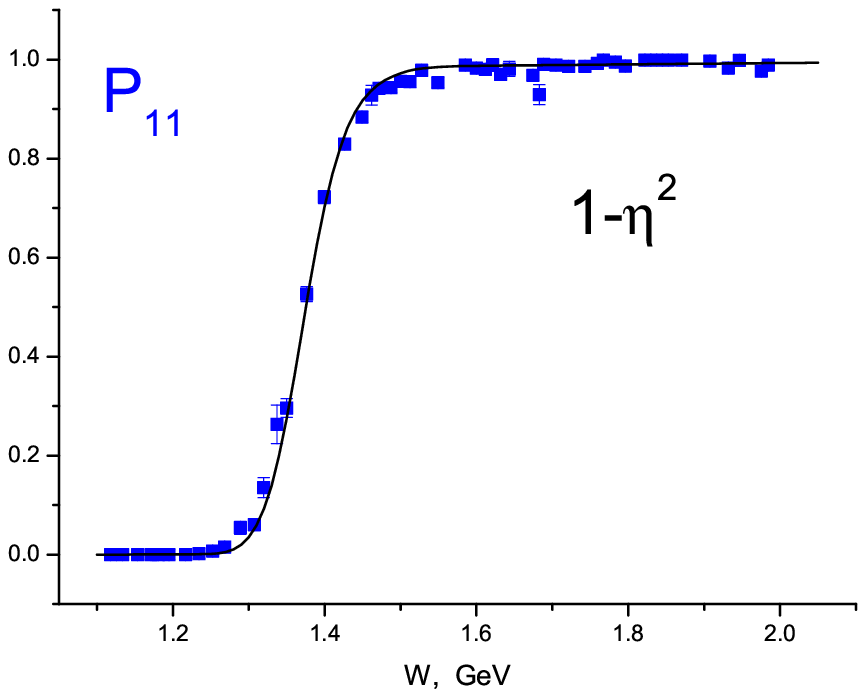}
  \caption{\label{fig:1} The results of fitting of $P_{11}$-wave of $\pi N$ scattering. Dots
    show results of PWA \cite{Arndt:2006bf}, solid lines represent our amplitudes
    \eqref{eq:K-matrix-pw-s-p}--\eqref{eq:K-bar-p-prim} in the presence of derivative in vertex
    \eqref{eq:10}--\eqref{eq:11}. $K$-matrix has only $p$-wave states. On the right side:
    $p$-wave inelasticity \cite{Arndt:2006bf}, the curve corresponds to lines on the left
    side. Partial wave normalization corresponds to \cite{Arndt:2006bf}: $\Im
    f=\abs{f}^{2}+(1-\eta^{2})/4$.}
\end{figure}

First of all, let us try to describe $S_{11}$ and $P_{11}$ waves separately. $p$-wave is
described rather well by our formulas with derivative in vertex \eqref{eq:10}--\eqref{eq:11},
see Fig.~\ref{fig:1}. In this case the $s$-wave states are missing in amplitudes, the $p$-wave
$K$-matrix has two positive energy poles.

Best-fit parameters corresponding to Fig.~\ref{fig:1} (in GeV units) are:
\begin{equation}\label{eq:12}
  \begin{split}
    m_{1} &=1.236\pm0.003, \quad g_{1,\pi}=7.93\pm0.07,\quad g_{1,\sigma} =8.47\pm0.11,\quad
    g_{1,\eta}=-2.90\pm0.20,\\
    m_{2} &= 1.504\pm0.001,\quad g_{2,\pi}=6.54\pm0.05,\quad g_{2,\sigma}=6.76\pm0.06,\quad
    g_{2,\eta}=5.0\;(\text{fixed}),\\
    m_{\sigma} &=0.3\;(\text{fixed}), \qquad \chi^{2}/\text{DOF}=273/95.
  \end{split}
\end{equation}
The use of vertices without derivative leads to impairment of quality of description:
$\chi^{2}>350$, again we need two poles with close masses.

\begin{figure}[!h]
  \centering
  \includegraphics{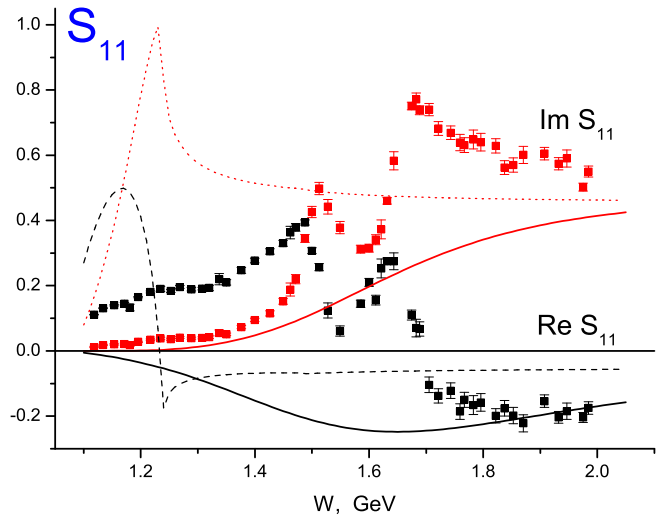}
  \caption{\label{fig:2} Background contribution to $s$-wave, generated by $p$-wave states,
    i.e. in this case $K$-matrix for $s$-wave \eqref{eq:K-bar-s-prim} has only negative energy
    poles. Solid lines represent variant with derivative in vertex (corresponding to curves on
    Fig.~\ref{fig:1}), dashed lines -- variant without derivative in vertex.}
\end{figure}

Both variants give a negative background contribution to $S_{11}$ wave, comparable in magnitude
with other contributions, as it seen on Fig.~\ref{fig:2}. Variant without derivative in vertex
gives a larger background contribution, rapidly changing near thresholds \footnote{Visible cusps
  in background contribution appear due to presence of large $p$-wave coupling constants,
  unnatural for $s$-wave, in this term. This is one of manifestations of OPF-mixing.}.  Of
course, we use rather rough approach -- effective $\sigma N$ channel can have different origin
in these waves. So, behavior of background contribution at low energy (especially without
derivative in vertex) is not well-defined. But it seems that description of $P_{11}$ partial
wave without derivative in vertices contradicts to data on $S_{11}$. On Fig.~\ref{fig:2} there
are shown some typical curves, there exist different variants with sharp behavior near
thresholds. The presence of derivative in a vertex suppresses the threshold region in background
contribution due to factor $(W-m_{N})^2$, but in resonance region this is rather large
contribution, see Fig.~\ref{fig:2}.

Attempt to describe partial wave $S_{11}$ without background contribution has no success: a
minimal variant of $K$-matrix with two positive energy poles don't allow to reach even
qualitative agreement with PWA.

\begin{figure}[!h]
  \centering
  \includegraphics{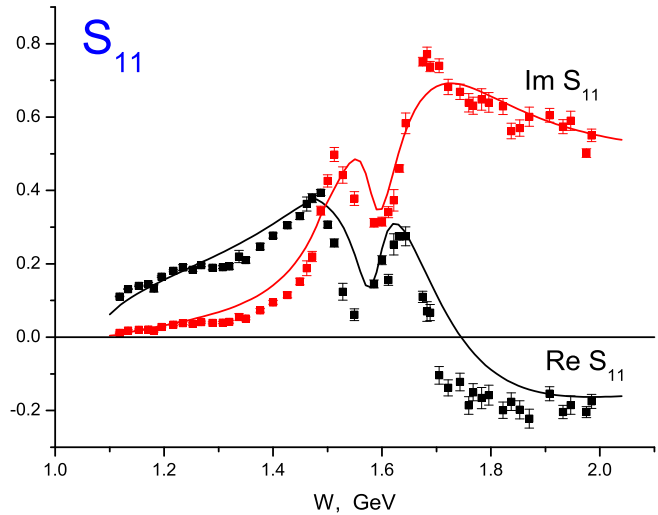}
  \caption{\label{fig:3} Results of $s$-wave fitting with fixed parameters for $p$-wave
    states. Parameters of $p$-wave correspond to curves on Fig.~\ref{fig:1}, $s$-wave contains
    two states with $K$-matrix masses $1.55$ and $1.75$ GeV.}
\end{figure}

As a next step, let us add the background contribution, arising from $p$-wave states (solid
lines on Fig.~\ref{fig:1}) with fixed parameters \eqref{eq:12}. One can see from
Fig.~\ref{fig:3} that quality of description is unsatisfactory in this case but double-peak
behavior is arisen in partial wave for the first time. It means that to describe $S_{11}$ wave
a background contribution is necessary and its value is close to solid line curves at
Fig.~\ref{fig:1}

\begin{figure}[!h]
  \centering
  \includegraphics[scale=0.8]{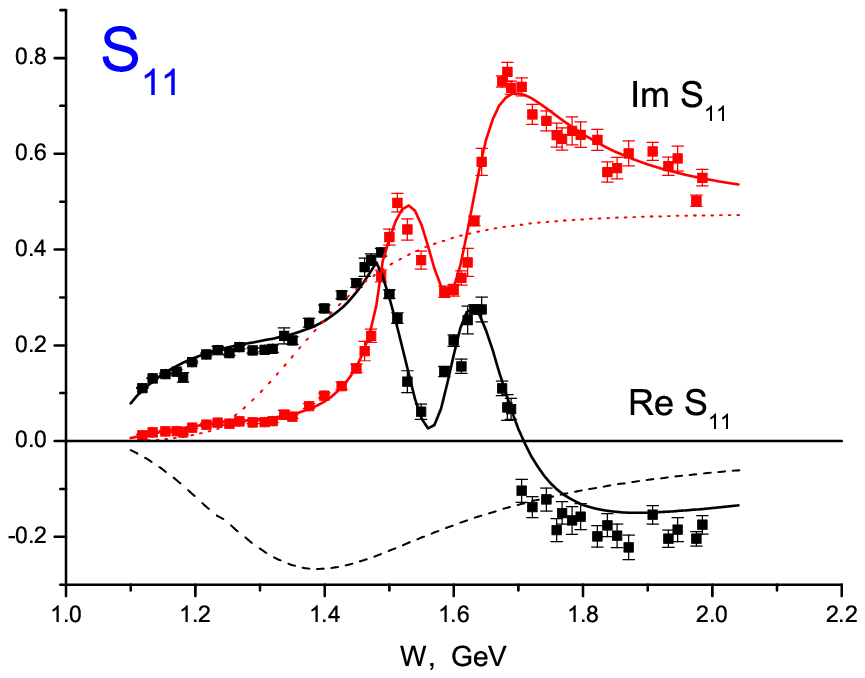}
  \includegraphics[scale=0.8]{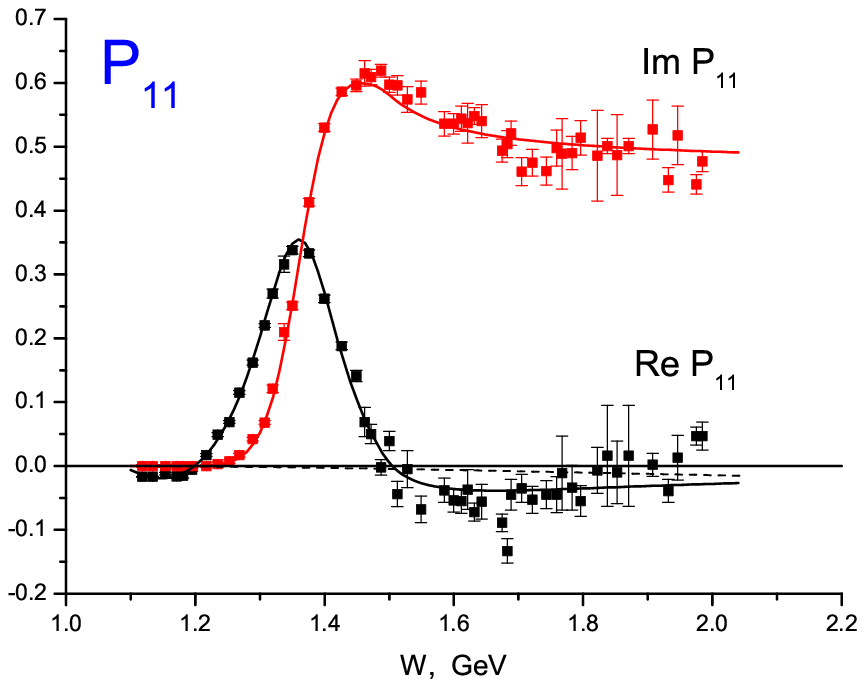}
  \caption{\label{fig:4} Result of joint fitting of $S_{11}$ and $P_{11}$-waves of $\pi N$
    scattering. $K$-matrices have two $s$-wave and two $p$-wave poles. Dashed lines show real
    and imaginary parts of (unitarized) background contribution. Imaginary part of background
    for $p$-wave is well below than real one and is not seen at figure.}
\end{figure}

Since the MacDowell symmetry connects two partial waves, it is naturally to perform the joint
analysis of $S_{11}$ and $P_{11}$ amplitudes, when resonance states in one wave generate
background in other and vice versa. In this case $K$-matrices
\eqref{eq:K-bar-s-prim}--\eqref{eq:K-bar-p-prim} have poles with both positive and negative
energy: we use two $s$-wave and two $p$-wave poles. This leads to noticeable improvement of
description, as it seen from Fig.~\ref{fig:4}; in this case $\chi^{2}/\text{DOF}=850/190$.

\begin{figure}[!h]
  \centering
  \includegraphics[scale=0.8]{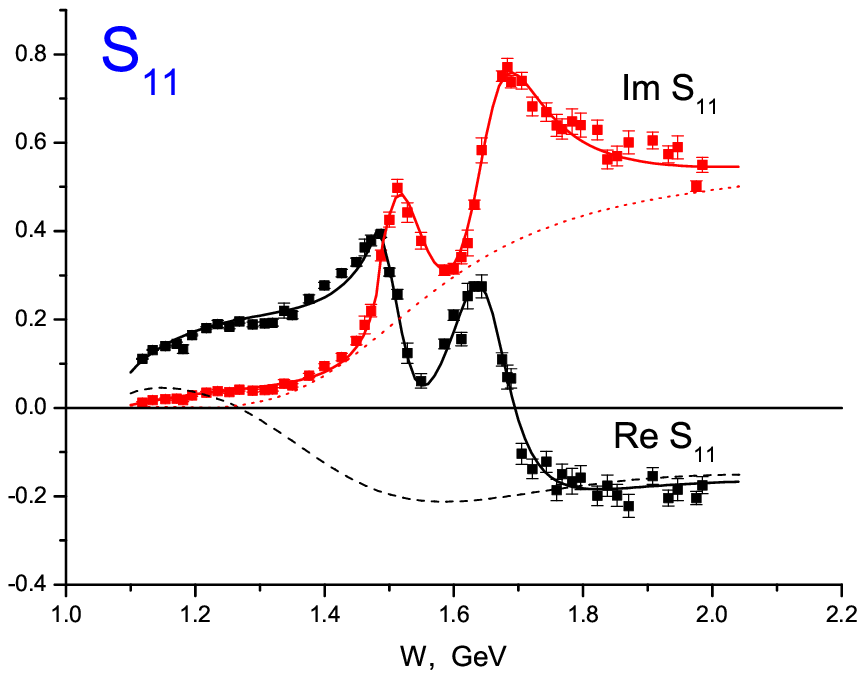}
  \includegraphics[scale=0.8]{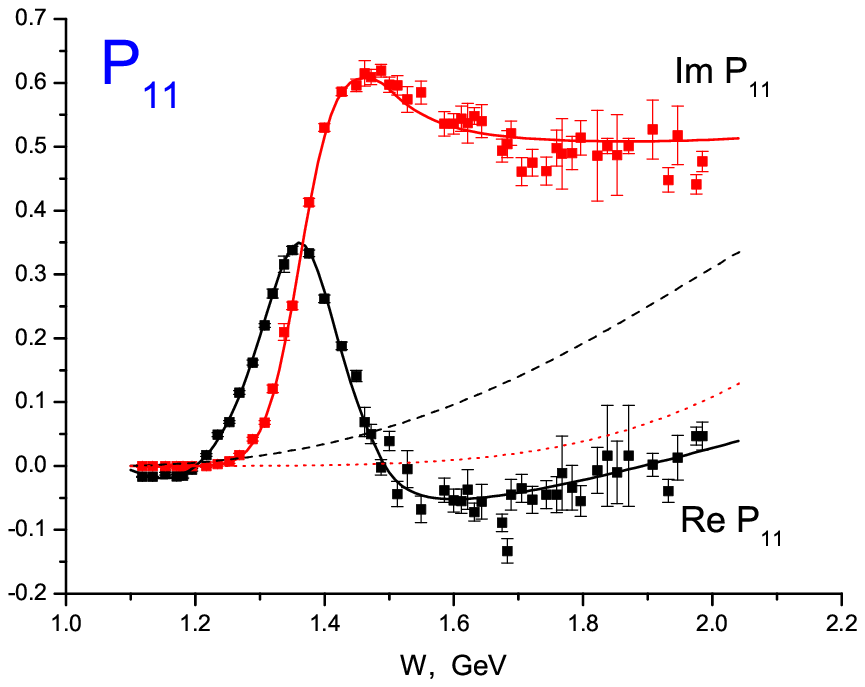}
  \caption{\label{fig:5} Result of joint fitting of $S_{11}$ and $P_{11}$ waves of $\pi N$
    scattering. $K$-matrix has two $s$- and two $p$-waves poles and background of form
    \eqref{eq:13}.}
\end{figure}

At last, background contributions can be generated not only by negative energy poles but by
other terms. We accounted it by adding to elastic amplitudes $\pi N\to\pi N$ a smooth
contributions of the form:
\begin{equation}
  \label{eq:13}
  \hat{K}_{s}^{B}=A+B(W-m_{N})^{2},\quad
  \hat{K}_{p}^{B}=A+B(W+m_{N})^{2},
\end{equation}
which do not violate the MacDowell symmetry property. Such terms correspond to pole
contributions with large masses in $s$- and $p$-waves. Results of joint description of two waves
are depicted at Fig.~\ref{fig:5}. Note that we have quite good description
$\chi^{2}/\text{DOF}=584/187$ and background contribution in $S_{11}$ is close to simplest
variant of Fig.~\ref{fig:2}.

So, the performed joint analysis of $S_{11}$ and $P_{11}$ partial waves demonstrates that
OPF-mixing gives rather marked effect in production of $1/2^{\pm}$ baryons.

In Table~\ref{tab:1} we present the values of pole masses and widths obtained by continuation of
our amplitudes to complex $W$ plane. As a whole, we see that our values for $m_{p}$,
$\Gamma_{p}$ are rather close to previously obtained. The only hint for disagreement is
appearance at some sheets of a stable pole $1/2^{+}$ with $m_{p}\approx 1500$ MeV instead of
generally accepted mass $m_{p}\approx 1365$ MeV. But this question as well as distribution of
poles over different Riemann sheets should be investigated in more correct multi-channel
approach, not with effective $\sigma N$ channel.

\begin{table}[!h]
  \centering
  \begin{tabular}{|c|c|c|}
    \hline
    \parbox[t]{0.25\linewidth}{Partial wave,\\[-1ex] PDG values\\[1ex]} &
    \parbox[t]{0.25\linewidth}{This work} & Some other works\\ \hline
    $S_{11}$, $1/2^{-}$  &              & \\
    N(1535) (1510, 70)  &  (1507, 87)  & (1502, 95), (1648, 80)   \cite{Arndt:2006bf}\\
    N(1650) (1655, 165) &  (1659, 149) & (1519, 129), (1669, 136) \cite{Doring:2009yv}\\ \hline
    $P_{11}$, $1/2^{+}$  &              & \\
    N(1440) (1365, 190) &  (1365, 194) & (1359, 162) \cite{Arndt:2006bf}  \\
                        &  (1500, 160) & (1385, 164) \cite{Hohler:1993xq} \\
                        &              & (1387, 147) \cite{Doring:2009yv} \\ \hline
  \end{tabular}
  \caption{\label{tab:1} Pole masses and widths $(M_{R},\Gamma_{R})$ extracted from poles position
    in the complex plane $W$: $W_{0}=M_{R}-\imath\Gamma_{R}/2$.}
\end{table}

\section{Conclusions}
\label{sec:concl}

In the present paper we investigated the manifestation of OPF-mixing in $\pi N$ partial waves $S_{11}$
and $P_{11}$, where baryons $1/2^{\pm}$, $I=1/2$ are produced. We found that the effect of mixing of
fermion fields with opposite parity can be readily realized in the framework of $K$-matrix
approach. It allows to have simple expressions for amplitudes in the case of any resonance states
and reaction channels. Note that $s$- and $p$-wave $K$-matrices,
\eqref{eq:K-bar-s-prim}--\eqref{eq:K-bar-p-prim}, have poles with both positive and negative
energy and are related with each other by $\hat{K}_{p}(W)=\hat{K}_{s}(-W)$.

The so constructed partial waves possess the well-known MacDowell symmetry that connects two partial
waves under substitution $W\to-W$. Up to now, this symmetry did not play any role in data
analysis since it connects physical and unphysical regions. However, taking OPF-mixing into
account, MacDowell symmetry leads to physical consequences: resonance in one partial wave gives
rise to background contribution in another and vice versa. This connection between two waves, as
in case of $3/2^{\pm}$ resonances \cite{Kaloshin:2010jj}, works mainly in one direction: it
generates large negative background in a wave with lower orbital momentum. So we come to idea of
joint analysis of two partial waves and it allows to get an additional information about
dynamics in higher $l$ wave. Such an example can be seen at Fig.~\ref{fig:2}, where two variants
of background in $S_{11}$ are depicted.

Our main purpose here was to see the effects of OPF-mixing in the amplitudes $S_{11}$, $P_{11}$
and to estimate their value. So, following \cite{Ceci:2006zw}, we have used simplified
three-channel formalism in which $\sigma N$ is some quasi-channel, imitating different $\pi\pi
N$ intermediate states. In spite of so rough approach we obtained rather good description of
$S_{11}$ and $P_{11}$ waves, comparable well with more comprehensive analyses
\cite{Nakamura:2010jn,Kamano:2010ud,Golli:2007sa,Paris:2010tz} with number of channels up to 6.
We suppose that OPF-mixing (or MacDowell symmetry) can be taken into account not only in
$K$-matrix formalism but in framework of more detailed dynamical multi-channel approach.

Note, that obtained pole positions not always coincide with the results of previous
analyses. For example, for $N(1440)$ state we found on most sheets a very stable pole with
$\Re{W}\approx1500$ MeV instead of "standard" value $\approx 1360$ MeV, see
Table~\ref{tab:1}. After various verifications we suppose that this is result of crudity of used
approximation (effective $\sigma N$ channel). But it is possible that here exists some
dependency on details of description and it needs more close investigation.

Summarizing, we found out that effect of a loop OPF-mixing is seen in PWA results as a
connection between partial waves $S_{11}$ and $P_{11}$. We assume that this connection may be of
interest as possibility to obtain additional information about $P_{11}$ wave and baryons
$1/2^{+}$.

\section{Bibliography}
\label{sec:biblio}

%\bibliographystyle{apsrev}
%\bibliography{biblio,roper}

\providecommand{\particle}[1]{\mathrm{#1}}

\end{document}